\documentclass[preprint,12pt]{elsarticle}

\usepackage{graphicx}
\usepackage{amssymb}

\usepackage{lineno}

\journal{}

\begin{document}

\begin{frontmatter}


\title{Criticality or supersymmetry breaking ?}



\author{\footnotesize Igor~V.~Ovchinnikov$^{1}$\footnote{email:igor.vlad.ovchinnikov@gmail.edu}, Wenyuan~Li$^{1}$, Yuquan~Sun$^{2}$, 
Robert~N.~Schwartz$^1$,
Andrew~E.~Hudson$^3$, Karlheinz~Meier$^4$, and Kang L. Wang$^1$\footnote{email:wang@seas.ucla.edu}}
\address{$^1$ Electrical and Computer Engineering Department, University of California at Los Angeles, Los Angeles, CA 90095 USA\\
$^2$ LMIB \& School of Mathematics and Systems Science, BeiHang University, Beijing 100191, China\\
$^3$ Department of Anesthesiology and Perioperative Medicine, David Geffen School of Medicine, UCLA, Los Angeles, CA USA\\
$^4$ Kirchhoff-Institute for Physics, Heidelberg University, Heidelberg, Germany}

\begin{abstract}
In many stochastic dynamical systems, ordinary chaotic behavior is preceded by a full-dimensional phase that exhibits 1/f-type power-spectra and/or scale-free statistics of (anti)instantons such as neuroavalanches, earthquakes, \emph{etc.} In contrast with the phenomenological concept of self-organized criticality, the recently developed approximation-free supersymmetric theory of stochastic differential equations, or stochastics, (STS) identifies this phase as the noise-induced chaos (N-phase), \emph{i.e.}, the phase where the topological supersymmetry pertaining to all stochastic dynamical systems is broken spontaneously by the condensation of the noise-induced (anti-)instantons. Here, we support this picture in the context of neurodynamics. We study a 1D chain of neuron-like elements and find that the dynamics in the N-phase is indeed featured by positive stochastic Lyapunov exponents and dominated by (anti)instantonic processes of (creation)annihilation of kinks and antikinks, which can be viewed as predecessors of boundaries of neuroavalanches. We also construct the phase diagram of emulated stochastic neurodynamics on Spikey neuromorphic hardware and demonstrate that the width of the N-phase vanishes in the deterministic limit in accordance with STS. As a first result of the application of STS to neurodynamics comes the conclusion that a conscious brain can reside only in the N-phase.
\end{abstract}


\end{frontmatter}

\section{Introduction}
It is well established by now that many stochastic dynamical systems close to the ''egde of chaos'' spontaneously exhibit features of long-range dynamical behavior such as 1/f power spectra or the scale-free statistics of instantonic or avalanche-like processes -- the phenomenon that can be found in all branches of modern science, including astrophysics \cite{XYZ}, finance \cite{Preis10052011}, geophysics \cite{GEO}, and evolutionary biology, \cite{BIO} collective human and animal \cite{Flocking_Birds,Flocking_Birds_2} behavior and many others including neurodynamics (ND) \cite{ChialvoLoh,Beggs02062004,Beggs03122003,EEG,Levina1}.

Such a spontaneous long-range dynamical behavior, which we loosely call dynamical complexity (DC), is a rather peculiar feature that calls for an explanation. One of the potential explanations of DC is ''criticality''. Namely, the very fact that DC is typically found on the border of chaos points to the possibility that the long-range features associated with DC may be attributed to the phase transition into chaos.

The phase transition picture of DC has one insoluble problem. On phase diagrams, which are spaces of externally controllable parameters that dynamical systems cannot change in the real time all by themselves, power-law correlators are manifested in the long-wavelength limit only by models that flow to unstable fixed points of the renormalization group (RG) flow. Such models occupy lower-dimensional boundaries between full-dimensional phases. On the contrary, DC occupies full-dimensional phases. Any model within a full-dimensional phase eventually flows to a stable fixed point (or other attractor) and such point represents a state with finite correlation length/time and consequently it cannot exhibit power-laws. Thus, DC occupying full-dimensional phases cannot in principle be explained within the paradigm of the traditional critical phenomena picture. 

To circumvent this problem with the criticality scenario for DC, it was proposed to believe that some stochastic dynamical systems have a mysterious tendency to fine tune themselves into the phase transition into chaos \footnote{Note that at the moment of this proposition, there existed no stochastic generalization of the concept of dynamical chaos.} -- the approach known as self-organized criticality (SOC) \cite{Bak1, Jensen_1, Zapperi_1,A_recent_review_on_SOC, PruessnerBook, XYZ, Frigg2003613,XYZ,PruessnerBook,PhysRevLett.78.4793, PhysRevLett.75.4528, ChristensenBook}, which is particularly popular in the neurodynamics (ND) community (for review see, \emph{e.g.}, Ref. \cite{10.3389/fnsys.2015.00022, 10.3389/fnsys.2014.00176,10.3389/fnsys.2014.00166, 10.3389/fnsys.2014.00154} and Refs. therein). After 25 years of the history of SOC, it is still unclear what SOC is exactly from a theoretical point of view (see Ref.\cite{A_recent_review_on_SOC} for a review on various interpretations of SOC) and whether ND in a healthy brain can be characterized as SOC (see, \emph{e.g.}, Refs. \cite{PhysRevLett.97.118102, 10.3389/fphys.2012.00302, 10.3389/fnsys.2014.00108, Zare201380, 10.3389/fnsys.2014.00151} and Refs. therein).

It was understood \cite{GoldstoneMode_1,GoldstoneMode_2} that a more rigorous theoretical picture of DC could be based on the Goldstone theorem stating that a spontaneous breakdown of a global continuous symmetry must lead to the long-range behavior in full-dimensional phases. The ubiquitous character of DC suggests that the Goldstone scenario for DC can work only if all (or at least most of) stochastic dynamical systems possess a common global continuous symmetry. The existence of such symmetry and the idea that its spontaneous breakdown is the theoretical essence of DC was proposed in Ref.\cite{Goldstone_Mechanism} and further work in this direction \cite{chaos_2} resulted in the formulation of the approximation-free supersymmetric theory of stochastics (STS) \cite{e18040108} viewing DC \cite{Resolution} as the spontaneous breakdown of topological supersymmetry (TS) (see, \emph{e.g.}, Refs. \cite{TFT_BOOK, Labastida_1989,Witten1,Witten2,Blau,Baulieu_1988,Baulieu_1989}) that all stochastic differential equations (SDEs) possess.

The presence of TS in all SDEs is the algebraic representation of the fact that any infinitesimally close points in phase space remain close during the (finite-time) evolution even in the presence of discontinuous noise \cite{Slavik}. In other words, TS is the preservation of the proximity of points in phase space during evolution. This interpretation makes it particularly clear that the spontaneous breakdown of TS must be viewed as the stochastic generalization of dynamical chaos.\cite{Kang,Max} Indeed, the spontaneous breakdown of TS must imply that initially close points may not be close anymore after infinitely long evolution, when the system is described by a non-supersymmetric ground state. In the deterministic limit, this is nothing other than the famous ''butterfly effect'' \cite{Lorenz}. 

\begin{figure}[t]
\centering
  \includegraphics[width=0.5 \linewidth]{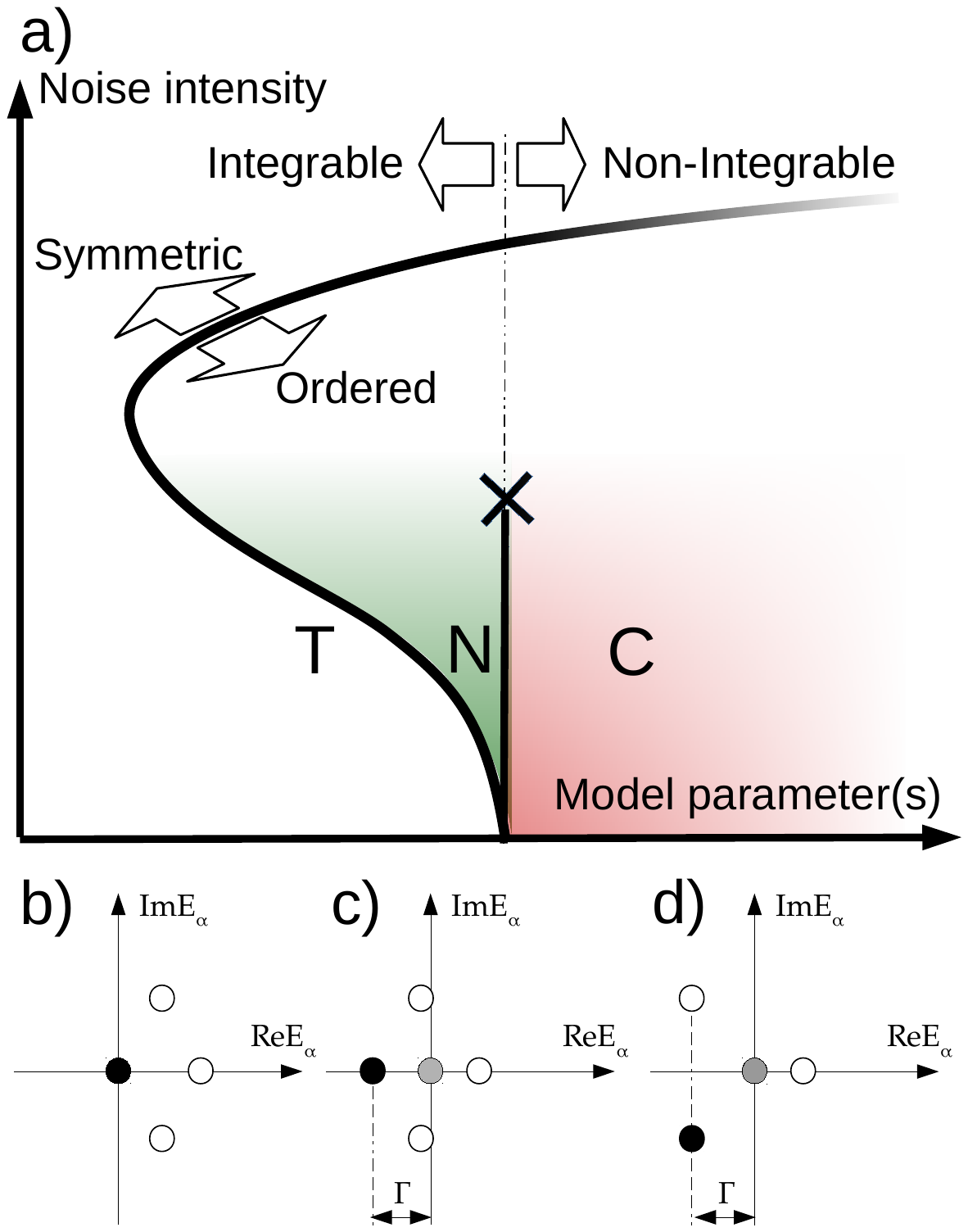}
\caption{{\bf (a)} The "border of chaos" as predicted by STS. Thick black curve separates systems with unbroken (symmetric) and spontaneously broken (ordered) topological supersymmetry (TS). Vertical straight line separates models with integrable and non-integrable FVF (laws of deterministic evolution). Accordingly, there are three phases in the weak noise limit: the phase of ordinary chaos (C, red) where TS is broken by non-integrability of FVF; the phase of noise-induced chaotic behavior (N, green) where TS is broken by condensation of noise-induced (anti)instantons; and the phase of ergodic dynamics or thermal equilibrium (T, white) with unbroken TS. In the deterministic limit, noise-induced antiinstantons disappear and the N-phase collapses into the T-C boundary. The termination of the N-C boundary by a cross at a certain noise intensity indicates that this boundary is transition-like only at weaker noises and at stronger noises it must be smeared out into a crossover. {\bf (b-d)} Three qualitatively different types of SEO spectra with unbroken (b) and broken (c-d) TS. Black dots are the ground states, which are the fastest growing eignenstates. Grey dots at the origin are supersymmetric states that are ground states only when TS is unbroken. In (d), there are two equally good candidates for the status of the ground state and the ground state is chosen between them by convention, similarly as it is done in quantum theory. $\Gamma$ is the real part of the ground state's eigenvalue providing a lower bound to the topological entropy of the model (see text).}
\label{Fig___1}
\end{figure}

On a more technical level, TS in the deterministic limit can be spontaneously broken only by the non-integrability of the flow vector field (FVF) \footnote{There are different types of integrability in mathematics. The integrability in the sense of dynamical systems means that the FVF has well-defined global unstable manifolds of all dimensions. When this is not true, the deterministic dynamical system is said to be non-integrable in the sense of dynamical systems of chaotic.}, which is one of the definitions of deterministic dynamical chaos \cite{Gilmore}. In the presence of noise, yet another mechanism of spontaneous breakdown of TS exists. This mechanism is known in the high-energy physics as the dynamical supersymmetry breaking \cite{Witten_Dyn_SUSY_Br} and, in the context of stochastic dynamics, it can be explained as the condensation of the noise-induced tunneling processes. Earthquakes, solar flares, neuroavalanches,	 \emph{etc}. are examples of such noise-induced tunneling processes and that phase with the (anti)instanton-induced breakdown of TS must be associated with DC. Furthermore, the noise-induced character of the tunneling processes requires that this phase disappear in the deterministic limit. In result, the phase of thermal equilibrium (unbroken TS, T-phase) and the phase of ordinary chaos (non-integrability, C-phase) can be separated (at weak noises) by the noise-induced chaos (TS broken by (anti)instantons, N-phase) as presented in Fig.\ref{Fig___1}a.

In this paper, we support this picture in two ways. First, we numerically investigate an overdamped stochastic 1D sine-Gordon model, which can be thought of as a coarse grained version of a 1D chain of neuron-like elements, and demonstrate that the stochastic Lyapunov exponents are positive in the N-phase. This confirms that TS is indeed spontaneously broken and the N-phase can indeed be identified as the noise-induced chaos. We also demonstrate that dynamics in the N-phase is dominated by the (anti)instantonic processes of the creation/annihilation of solitonic configurations (kinks-antikink pairs). Second, using 1/f noise as the experimental signature of the spontaneous breakdown of TS, we construct the rudimentary phase diagram of ND using emulation on \emph{Spikey} neuromorphic hardware.\cite{Karlheinz1,Karlheinz2} Our experimental results reconfirm that the width of the N-phase vanishes in the deterministic limit. As a first non-trivial result from the application of STS to ND, we present arguments supporting the conclusion that a healthy brain can only reside in the N-phase.

The structure of the paper is as follows. In Sec.\ref{STS_explanation}, we briefly present the key elements of STS. In Sec.\ref{sine_Gordon}, we introduce an overdamped stochastic 1D sine-Gordon model and present our results proving that the N-phase must indeed be recognized as the noise-induced chaos with instanton-induced TS breaking. In Sec.\ref{Emulation}, we present our results on the emulation of stochastic ND using \emph{Spikey} neuromorphic hardware reconfirming that the width on the N-phase vanishes in the deterministic limit. In Sec.\ref{Conclud}, we conclude with a brief discussion of potentially fruitful directions for further investigation.

\section{Key elements of STS}
\label{STS_explanation}


In this section, we would like to briefly discuss the key elements of STS, while the details on this theory can be found in Ref.\cite{e18040108}. The main object of interest is the general form of a SDE that can be given as,
\begin{eqnarray}
\dot x(t) = F(x(t)) + (2\Theta)^{1/2}e_a(x(t))\xi^a(t) \equiv  {\tilde F}(t).\label{SDE}
\end{eqnarray}
Here and in the following, summation is assumed over repeated indices, $x\in X$ is a point in the phase space, $X$, which is a smooth topological manifold, $F\in TX$ is the FVF representing the deterministic equations of motion on the tangent space $TX$, $e_a\in TX$ is a set of vector fields describing how noise is coupled to the system, $\Theta$ is the intensity or the temperature of the noise, $\tilde F$ is introduced for later convenience, and $\xi^a\in \mathbb{R}^{Dim X}$ is a set of noise variables that are assumed Gaussian white with the probability of configurations given by,
\begin{eqnarray}
P(\xi)\propto e^{- \int_{t'}^td\tau \xi^2(\tau)/2}.\label{prob}
\end{eqnarray}

\subsection{From SDE to topological field theory}


The Parisi-Sourlas approach \cite{ParSour,ZinnJustin1986} provides pathintegral representation to Langevin SDEs, which are almost exclusively studied in the context of supersymmetry and stochastics (see, \emph{e.g.}, Ref.\cite{Drummond_Horgan_2012}). STS can be looked upon as the generalization of the Parisi-Sourlas approach to the SDEs of arbitrary form. More specifically, one can construct the following functional,
\begin{eqnarray}
{  W} = \left\langle \int_{p.b.c.} Dx\prod_\tau \delta(\dot x(\tau) - { \tilde F}(\tau)){Det}\frac{\delta(\dot x - { \tilde F} )}{\delta x}\right\rangle.\label{PathIntWittenIndex}
\end{eqnarray}
The pathintegral here is over closed paths in $X$ as indicated by the subscript p.b.c. denoting periodic boundary conditions, $x(t)=x(t')$, the infinite-dimensional determinant is the Jacobian of the preceding $\delta$-functional, which, in turn, limits the functional integration to summation over the solutions of the SDE. The angled brackets denote stochastic averaging over the noise configurations,
\begin{eqnarray}
\langle A(\xi) \rangle = \left(\int D \xi P(\xi)\right)^{-1}\int D \xi A(\xi) P(\xi), 
\end{eqnarray}
with $A(\xi)$ being an arbitrary functional of $\xi(\tau)$ and the probability $P$ given in Eq.(\ref{prob}).

To exponentiate the bosonic $\delta$-functional and its determinant in Eq.~(\ref{PathIntWittenIndex}) one can use the standard technique of introducing additional fields: the Lagrange multiplier, $B_i$, and the pair of Faddeev-Popov ghosts, $\chi^i$ and $\bar\chi_i$, with $1\le i\le Dim X$. This procedure leads from Eq.~(\ref{PathIntWittenIndex}) to,
\begin{eqnarray}
{  W} = \left\langle \int_{p.b.c.} D\Phi e^{\{ Q, i\int_{t'}^t d\tau \bar \chi(\tau)(\dot x(\tau)-\tilde F(\tau))\} } \right\rangle,\label{Witten_Path_1}
\end{eqnarray}
and after integrating out the noise
\begin{eqnarray}
{  W} = \int_{p.b.c.} D\Phi \; e^{\{{  Q},\Psi\}}.\label{Witten_Path}
\end{eqnarray}
Here, $\Phi=(x,B,\chi,\bar\chi)$ denotes the collection of all the fields, periodic boundary conditions are assumed, $\Phi(t)=\Phi(t')$, and the operator of the Becchi-Rouet-Stora-Tyutin (BRST) symmetry is defined as
\begin{eqnarray}
\{ {  Q}, A\} = \int_{t'}^t d\tau \left(\chi^i(\tau)\frac{\delta }{\delta x^i(\tau)} + B_i(\tau)\frac{\delta }{\delta \bar\chi_i(\tau)}\right) A, \label{Q_pathint}
\end{eqnarray}
with $A$ being an arbitrary functional of $\Phi$, and
$\Psi = \int^{t}_{t'} \left(i \bar\chi_j(\tau) \dot x^j(\tau) - \bar d(\Phi(\tau))\right)d\tau$ being the so-called gauge fermion with
\begin{eqnarray} 
\bar d(\Phi) = i\bar\chi_j\left(F^j - \Theta e_a^j \{{  Q}, i\bar \chi_ke_a^k\}\right), \label{Current_Op_Path}
\end{eqnarray}
which can be loosely identified as a probability current operator. The Guassian integration over $\xi$ in Eq.(\ref{Witten_Path_1}) leads to a term in the Lagrnagian $\propto \{Q, i\bar\chi_je_a^j \}\{Q, i\bar \chi_ke_a^k\}$ which is $Q$-exact, due to the nilpotency of BRST symmetry. Namely, $\{ Q, \{Q , A \} \}=0, \forall A $ implies 
\begin{eqnarray}
  \{Q, i\bar\chi_je_a^j \}\{Q, i\bar \chi_ke_a^k\} 
= \{Q, i\bar\chi_je_a^j \{Q, i\bar \chi_ke_a^k\}\}.\label{Q_Exact}
\end{eqnarray}
This is the reason why integrating out the noise in Eq.(\ref{Witten_Path}) leaves the action of the model $Q$-exact. 

In gauge theories, $Q$-exact pieces in actions are essentially gauge fixing tools. It is typically said that BRST symmetry, as a symmetry transformation, generates (fermionic versions of) gauge transformations and its overall effect on the pathintegral representation of the theory is to fix the gauge. This interpretation applies to Eq.(\ref{Witten_Path}), too -- the BRST generates all possible (fermionic) deformations of the path, $\delta x(\tau)=\chi(\tau)$, and out of all the possible closed paths in $X$, it leaves only the solutions of SDE (\ref{SDE}) as can be best seen from Eq.(\ref{PathIntWittenIndex}).

Identification of the stochastic quantization procedure as gauge fixing has a hidden danger. Indeed, in gauge theories, $Q$-exact pieces, as the one in Eq.(\ref{Witten_Path}), appear in generating functionals (or partition functions) that can be used to calculate various observables including responses of the model to external perturbations. Therefore, it may be tempting to believe that Eq.(\ref{Witten_Path}) is the generating functional of the SDE. This assumption is a common mistake. The point is that Eq.(\ref{Witten_Path}) is not a generating functional. It is actually a topological invariant, which is not responsive to external perturbations. Any response correlator calculated within it vanishes. 

This brings us to the topological field theory \cite{TFT_BOOK} side of the story of STS. Namely, the BRST symmetry in Eq.(\ref{Q_pathint}) can be also rightfully identified as a TS; the integrand in Eq.(\ref{Witten_Path}) as a member of the mathematical construction called the Mathai-Quillen class \cite{Blau}; and the model itself as a member of the Witten-type topological or cohomological field theories \cite{Witten2,Witten1}, the models whose actions look like gauge fixing of an empty theory and that have intrinsic connection to Morse theory \cite{Witten_1982,Labastida_1989}.


\subsection{Operator representation}


Object $W$ in Eq.(\ref{Witten_Path}) is the famous Witten index. It is of topological character that can be established using the operator representation of the theory. The later is achieved in a standard matter. First, one rewrites 
\begin{eqnarray}
{  W} = \int_{p.b.c.} D\Phi \; e^{\int^t_{t'} d\tau (i \dot x^j B_j+ i \dot{\chi^j}\bar{\chi}_j - H(\Phi(\tau))},\label{Witten_Pre_Oper}
\end{eqnarray}
where
\begin{eqnarray}
H(\Phi) = \left\{  {Q}, \bar{d}(\Phi) \right \},\label{SEO_Path}
\end{eqnarray}
with $\bar{d}(\Phi)$ defined in Eq.(\ref{Current_Op_Path}). It is clear from Eq.(\ref{Witten_Pre_Oper}) that $B$ and $\bar\chi$ are momenta fields that must (anti-)commute with $x$ and $\chi$ in the operator representation,
\begin{eqnarray}
[i\hat B_j , \hat x^k] = [i\hat {\bar \chi}_j , \hat \chi^k]  = \delta^k_j, 
\end{eqnarray}
where square brackets denote bi-graded commutator, which is an anticommutator when both operators are fermionic and a commutator otherwise. 

Now, working in the representation where operators $\hat x^i$ and $\hat \chi^i$ are diagonal ($\hat x^i\equiv x^i, \; \hat \chi^i\equiv \chi^i$) so that $i\hat B_j = \partial/\partial x^j$ and $i\hat {\bar \chi}_j = \partial/\partial \chi^j$, and using Wick symmetrization rule in order to bypass the operator ordering ambiguity, \footnote{It can be shown \cite{e18040108} that Wick symmetrization rule is equivalent to the Stratonovich interpretation of SDEs. It can also be shown that the most natural mathematical interpretation of the finite-time stochastic evolution operator, $e^{-(t-t')\hat H}$, is the stochastically averaged pullback induced by the SDE-defined diffeomorphisms of the phase space. This interpretation unambiguously resolves the Ito-Stratonovich dilemma \cite{ISD_1,ISD_2,ISD_3} in favor of the Stratonovich approach.} one finds the following operator version of Eq.(\ref{Witten_Pre_Oper}),
\begin{eqnarray}
{  W} = Tr (-1)^{\hat k} e^{-(t-t')\hat H}, \label{Witten_Oper}
\end{eqnarray}
with,  
\begin{eqnarray}
\hat k = \chi ^j\frac\partial{\partial \chi^j},
\end{eqnarray}
being the fermion number operator and the (infinitesimal) stochastic evolution operator (SEO) defined as,
\begin{eqnarray}
\hat H = \hat {  L}_F - \Theta \hat {  L}_{e_a}\hat {  L}_{e_a},
\end{eqnarray}
where ${  L}$'s are the Lie derivatives along the subscript vector fields. The number of fermions commutes with the SEO,
\begin{eqnarray}
[\hat k, \hat H] = 0,\label{k_H}
\end{eqnarray}
which implies that stochastic evolution preserves the number of fermions.

Recalling the Cartan formula, 
\begin{eqnarray}
\hat{  L}_{F} = [\hat d, \frac{\partial}{\partial \chi^j} F^j ],
\end{eqnarray}
where 
\begin{eqnarray}
\hat d = \chi^j \frac\partial{\partial x^j}, 
\end{eqnarray}
is the exterior derivative, one can recast the SEO into the explicitly supersymmetric form,
\begin{eqnarray}
\hat H = [\hat d, \hat {\bar d}],\label{SUSY}
\end{eqnarray}
where 
\begin{eqnarray}
\hat {\bar d} = \frac\partial{\partial \chi^j} \left(F^j - \Theta e_a^j\hat {  L}_{e_a}\right).\label{Current_Op}
\end{eqnarray}
Eqs.(\ref{SUSY}) and (\ref{Current_Op}) are the operator versions of Eqs.(\ref{SEO_Path}) and (\ref{Current_Op_Path}) respectively. This understanding also suggests that the (bi-graded) commutation with $\hat d$ is the operator version of the TS: $\{ {Q},A(\Phi)\} \to [\hat d, A(\hat \Phi)]$. The exterior derivative is indeed a (super)symmetry of the model because it is commutative with SEO,
\begin{eqnarray}
[\hat d, \hat H]=0.
\end{eqnarray}
This follows from $\hat d$-exactness of $\hat H$ in Eq.(\ref{SUSY}) and nilpotency of the exterior derivative, $\hat d^2=0$, which implies, in particular, that $[\hat d, [\hat d,\hat A]]=0, \forall \hat A$ (c.f.  Eq.(\ref{Q_Exact})).


\subsection{Eigensystem}


The SEO is a real operator. Therefore, it is pseudo-Hermitian \cite{Mos023}. As a pseudo-Hermitian operator, SEO has a complete bi-orthogonal eigensystem with the left (bras) and right (kets) eigenstates such that,
\begin{eqnarray}
\langle n| \hat H = \langle n |  {E}_n, \; \; \hat H |n \rangle =  {E}_n | n\rangle,\label{BrasKets}\\
\langle n | k \rangle = \delta_{nk}, \; \; \sum\nolimits_{n} |n\rangle\langle n | = \hat 1_{\Omega},
\end{eqnarray}
Here, $\Omega$ is the Hilbert space of the model, which is the exterior algebra of $X$. Namely, in the representation where $\hat x$ and $\hat \chi$ are diagonal, a wavefucntion is a function of the position on the phase space, $x\in X$, and a Grassmann variable, $\chi$. The later can be viewed \cite{Witten_1982} as differentials of differential forms, $\chi^l\chi^m = - \chi^m\chi^l \sim - dx^m\wedge dx^l=dx^l\wedge dx^m$. A general wavefunction can then be given as,
\begin{eqnarray}
\psi(x\chi) = \sum\nolimits_{k}\psi^{(k)}(x\chi),
\end{eqnarray}
where
\begin{eqnarray}
\psi^{(k)}(x) &=& k!^{-1}\psi_{i_1..i_k}(x)\chi^i...\chi^k  
\nonumber \\
&\equiv& k!^{-1}\psi^{(k)}_{i_1...i_k}(x) dx^{i_1} \wedge ...\wedge dx^{i_k} \in \Omega^k(x),
\end{eqnarray}
are differential forms on $X$ of degree $k$. 

It must be stressed that unlike in quantum mechanics with Hermitian evolution operator, the relation between bras and kets in Eq.(\ref{BrasKets}) is not trivial due to the pseudo-Hermiticity of SEO.

The presence of TS divides all eigenstates into two groups. The majority of states belong to the first group of non-supersymmetric doublets or pairs of eigenstates,
\begin{eqnarray}
|\alpha\rangle  \; \textmd{and} \;  |\alpha'\rangle = \hat d|\alpha\rangle.\label{NonSusyDoublet}
\end{eqnarray}
Eigenstates within each doublet have the same eigenvalue. Indeed,
\begin{eqnarray}
 {E}_{\alpha} |\alpha\rangle = \hat H|\alpha \rangle \to    {E}_{\alpha} \hat d |\alpha\rangle = \hat d \hat H|\alpha \rangle = \hat H (\hat d|\alpha \rangle). 
\end{eqnarray}
Note that Eq.(\ref{k_H}) implies that the operator $\hat k$ can be diagonalized together with the SEO,
\begin{eqnarray}
\hat k |n\rangle = k_n | n \rangle. 
\end{eqnarray}
In other words, each eigenstate has a well-defined number of fermions, $k_n$. Further, since the exterior derivative raises the number of fermions by one, the following is true, $k_{\alpha'}=k_\alpha+1$ for supersymmetric doublets in Eq.(\ref{NonSusyDoublet}). This implies that the contribution from pairs of non-supersymmetric states cancel out in Eq.(\ref{Witten_Oper}). $W$ receives a contribution only from supersymmetric singlets -- the eigenstates that obey,
\begin{eqnarray}
\hat d |\theta\rangle=0, \;|\theta\rangle\ne \hat d |x\rangle, \forall |x\rangle. \label{susy_states}
\end{eqnarray}
The key property of supersymmetric eigenstates is the vanishing expectation values for all $\hat d$-exact operators, $\langle \theta | [\hat d, \hat A] |\theta \rangle, \forall \hat A$. The SEO is a $\hat d$-exact operator. Therefore, the $\theta$'s have strictly zero eigenvalue. In result,
\begin{eqnarray}
{  W} = \sum\nolimits_{\theta} (-1)^{k_{\theta}}. \label{Euler}
\end{eqnarray}
It must be pointed out that condition (\ref{susy_states}) is essentially the requirement for a state to be non-trivial in de Rahm cohonology. For compact $X$, each de Rahm cohomology must provide one supersymmetric eigenstate -- otherwise the eigensystem of $\hat H$ would be incomplete. Therefore, Eq.(\ref{Euler}) equals the Euler characteristic of $X$. This completes the demonstration of the topological character of the Witten index, $W$.


\subsection{Dynamical partition function, supersymmetry breaking, and chaos}


The alternating sign factor in Eqs.(\ref{Witten_Oper}) and (\ref{Euler}) appears due to the unconventional periodic boundary conditions for fermionic fields in Eq.(\ref{Witten_Path}) -- normally, one would expect antiperiodic boundary conditions for fermionic fields. As we already said before, $W$ is a not a generating functional and/or dynamical partition function (DPF) of the model. The later can be obtained from Eq.(\ref{Witten_Path}) by switching to the anti-periodic boundary conditions (a.p.b.c.) for the fermionic fields. This removes the alternating sign factor in the operator representation of the DPF,
\begin{eqnarray}
{  Z} = \int_{a.p.b.c.} D\Phi \;e^{\{{  Q},\Psi\}}  = Tr e^{-(t-t')\hat H}.
\end{eqnarray}

In the long time limit, $t-t'\to\infty$, only eigenstates, $p$, with the lowest real part of their eigenvalues, $Re {E}_p=-\Gamma_g$, 
\begin{eqnarray}
\Gamma_g=-\min_n Re {E}_n \ge 0,
\end{eqnarray} 
contribute into the DPF,
\begin{eqnarray}
\left. {Z}\right|_{t-t'\to\infty} \propto e^{(t-t')\Gamma_g}\cos (t-t') Im  {E}_g.\label{Growth}
\end{eqnarray}
These eigenstates can be identified as physical states, whereas one of the physical states can be declared the ground state of the model. Among the three possible types of SEO spectra given in Fig.\ref{Fig___1}b-d, the two spectra with non-zero $\Gamma_g>0$ have TS broken spontaneously because the corresponding ground state have non-zero eigenvalue and thus it is non-supersymmetric. 

We would also like to note that both types of supersymmetry broken SEO spectra are realizable. This follows from the recently established relation \cite{Torsten} between STS and the astrophysical phenomenon of the kinematic dynamo \cite{Li20108666,KD_1,KD_2,Old_KD_Results,Ott1998} and the fact that the evolution operator in the theory of the kinematic dynamo is known to have both kinds of spectra with the real and the complex ground states. 

Unlike the Witten index, the DPF contains vital information about the dynamical properties of the model. In particular, it can also be shown that for a wide class of models and in the long-time limit, DPF provides a lower bound for the stochastically averaged number of periodic solutions of SDE,
\begin{eqnarray}
\left. {Z}\right|_{t-t'\to\infty} \le \langle \#({periodic\; solutions})\rangle = e^{(t-t')S},\label{Top_Entrop}
\end{eqnarray}
where $S$ is the stochastic generalization of the topological entropy of dynamical system \cite{Adler} -- the measure of "complexity" of dynamics. Positiveness of $S$ is considered the key feature of dynamical chaos. From Eqs.(\ref{Top_Entrop}) and (\ref{Growth}) it follows that $\Gamma_g$ provides the lower bound for $S$,
\begin{eqnarray}
\Gamma_g \le S.
\end{eqnarray}
Therefore, when TS is broken spontaneously, $\Gamma_g$ together with $S$ are positive and it immediately follows that the phenomenon of the spontaneous breakdown of TS must be associated with the stochastic generalization of dynamical chaos \cite{Kang}. 

Yet another proof that spontaneous TS symmetry breaking must be indeed identified as the stochastic generalization of dynamical chaos is the Goldstone theorem stating that a model with spontaneously broken TS must exhibit a long-range dynamical behavior, which can be interpreted as the emergent dynamical memory of initial conditions widely known as the ''butterfly effect'' of chaos. 

In numerical experiments, this infinitely-long memory can be characterized by positive Lyapunov exponents and we will use this well-established approach in Sec.\ref{sine_Gordon}. In real experiments, the spontaneous TS breaking can reveal itself via emergent 1/f-type power-spectra that we will use in Sec.\ref{Emulation} as the signature of the TS breaking in emulated ND.

\subsection{The phase diagram}

The phase diagram in Fig.\ref{Fig___1} can now be understood as follows. In the deterministic limit, the spontaneous TS breaking is equivalent to the concept of deterministic chaos. Further, in the limit of strong noise, $\Theta\to\infty$, the SEO is dominated by the diffusion Laplacian, $\hat L_{e_a}\hat L_{e_a}$, and diffusion alone should not break TS.\footnote{This is certainly true for those models where the diffusion Laplacian equals the Hodge-Laplacian that has real non-negative eigenvalues.} In other words, the TS will be eventually restored as one rises the intensity of the noise. The above two observations leave only two possible forms of the ''border of chaos''. One is that the TS broken phase monotonously shrinks with the increase of $\Theta$ until it disappears completely. The other, more interesting possibility is presented in Fig.\ref{Fig___1}a, where the TS broken phase first grows with $\Theta$ giving rise to the N-phase with integrable FVF and spontaneously broken TS. 

An important question is the mechanism of spontaneous TS breaking in the N-phase. In deterministic limit, the TS is broken by non-integrability of the FVF, which is one of the definitions of deterministic chaos \cite{Gilmore}. In the N-phase, the FVF is integrable so that some other mechanism must be responsible for the TS breakdown. There are two known candidates. The first is the anomaly or perturbative corrections that in the context of stochastic dynamics represent fluctuations due to noise. This mechanism of TS breaking is very unlikely due to what is known as non-renormalization theorems. Therefore, it must be the other remaining known mechanism that must be responsible for the TS breaking in the N-phase. This mechanism is the condensation of (anti)instantonic processes. This mechanism is known in the high-energy physics as dynamical supersymmetry breaking. \cite{Witten_Dyn_SUSY_Br} In the case of stochastic dynamics, the antiinstantoic processes are essentially the noise-induced tunneling processes between, \emph{e.g.}, different attractors of the FVF. The nose-induced character of the tunneling processes explains why the N-phase disappears in the deterministic limit.

Yet another important issue is the N-C boundary. At low temperatures, when an external observer can tell one tunneling process from another, the N-C boundary must behave as a phase transition or, rather, as the sharp onset of ordinary chaotic behavior. At higher temperatures, the N-C boundary must be smeared into a crossover because the tunneling processes must overlap in time (and space) and an external observer will not be able to tell one tunneling process from the other. The disappearance of the sharp N-C boundary does not contradict any symmetry-based argument because the N-C boundary is not the TS symmetry breaking in the first place.

\begin{figure}[t]
\centering
\includegraphics[width=0.6 \linewidth]{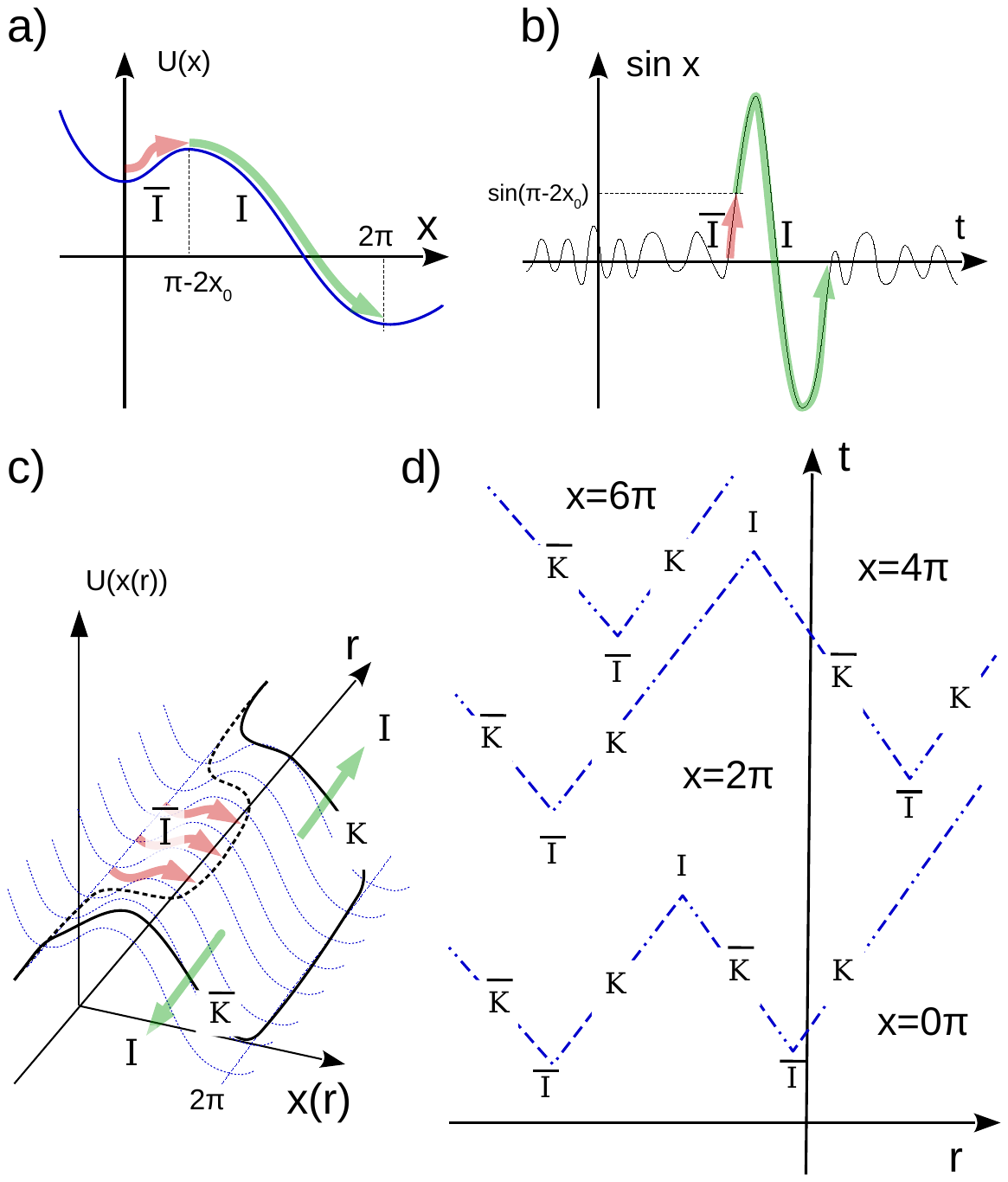}
\caption{{\bf (a)} Local potential function (\ref{potential}) and the process of tunneling from $x=0$ to $x=2\pi\sim 0$, which is a combination of an antiinstanton, $\bar I$, leading from $x=0$ to the unstable critical point $x=\pi-2x_0$, and an instanton, $I$, leading further to $x=2\pi$. {\bf (b)} In certain coordinates, the noise-induced tunneling process is spike-like, similarly to that in real neurons. {\bf (c)} In 1D chain of neuron-like elements, the noise-induced antiinstanton becomes a process of creation of a pair of solitions, a kink ($K$) and antikink ($\bar K$). Free solitons travel with constant velocity in opposite directions. The process of creation of $K-\bar K$ pair with subsequent propagation of solitions in different directions is this model's counterpart of neuroavalanches. {\bf (d)} Qualitative representation (a contour plot for $x(tr)$) of the dynamics in the N-phase under the condition of condensation of $\bar I$'s and instantons ($I$), the processes of creation and annihilation of the $K$-$\bar K$ pairs.}
\label{Fig___2}
\end{figure}

\section{1D chain of stochastic neuron-like elements}
\label{sine_Gordon}

Our goal in this section is twofold. First, we would like to demonstrate numerically that the N-phase can indeed be identified as a noise-induced chaotic dynamics. This goal will be achieved by revealing positiveness of the stochastic Lyapunov exponents in the N-phase. Second, we would also like to demonstrate that dynamics in the N-phase is indeed dominated by (anti)instantons. Instantons are easily recognizable when they have a well-defined spatio-temporal structure (creation/annihilation of kink-antikink pairs, see below), which is realized in spatially extended models on lattices. Guided by this understanding and by the additional intension to make connection to ND in the next section, here we study neuron-like elements on the simplest lattice - a 1D chain.

Each neuron-like element is described by a dynamical variable $x\in \mathbb{S}^1$. In isolation and in the absence of noise, its dynamics is governed by the following ordinary differential equation,
\begin{eqnarray}
\dot x(t)  = \alpha - \sin(x(t) + x_0).\label{ODE_1}
\end{eqnarray}
Here, constant $x_0 = \sin^{-1}\alpha$ is introduced for convenience so that $x=x_s=0$ is the stable critical point (for $\alpha<1$ see below). There is also an unstable critical point at $x = x_u = \pi - 2x_0$. Locally, the FVF can be viewed as a gradient of the potential function,
\begin{eqnarray}
\alpha - \sin(x + x_0) = - \partial U(x)/\partial x, \; U(x) = - \alpha x - \cos(x+x_0),\label{potential}
\end{eqnarray}
as displayed in Fig.\ref{Fig___2}a. This figure also depicts the fundamental tunneling process of the model. This process has two stages called antiinstanton ($x_s\to x_u$) and instanton ($x_u \to x_s = 2\pi \sim x_s$). In general, antiinstantons are processes of going against the FVF, \emph{i.e.}, the r.h.s of Eq.(\ref{ODE_1}). Therefore, antiinstantons can only happen as a result of the influence of noise, unlike instantons that exist even in the deterministic limit.  

The above neuron-like elements are now arranged into a 1D chain with the coupling between nearest neighbors. The resulting model can be coarse-grained into a continuous-space model defined by the following SDE,
\begin{eqnarray}
\partial_t x(rt)  = \alpha - \frac{\delta V_{sG}}{\delta x}(rt)+ (2\Theta)^{1/2}\xi(rt),\label{SDE_1}
\end{eqnarray}
where $r$ is a spatial dimension, the noise $\xi$ is Gaussian white,
\begin{eqnarray}
\langle \xi(rt) \rangle = 0, \langle \xi(rt)\xi(r't') \rangle = \delta(t-t')\delta(r-r'), 
\end{eqnarray}
and $\Theta$ is the noise intensity. The sine-Gordon potential in Eq.(\ref{SDE_1}) is defined as,     
\begin{eqnarray}
\frac{\delta V_{sG}}{\delta x}(rt) = -\partial_r^2 x(rt) + \sin(x(rt) + x_0), \\
V_{sG}(x) = \int dr \left((\partial_r x )^2/2 - \cos(x - x_0)\right).
\end{eqnarray}
In addition to being a coarse-grained version of the 1D chain of neuron-like elements, the above model can also be viewed as the overdamped limit of a stochastic sine-Gordon equation \cite{sine_Gordon_Kink_Bath,Classics,Josephson_general} or Frenkel-Kontorova equation \cite{Review_Frenkel_Kontorova}. Furthermore, in the theory of 1D chains of Josephson junctions, the model describes the temporal evolution of the voltage.\cite{PhysRevB.54.1234}

\begin{figure}[t]
\centering
\includegraphics[width=0.6 \linewidth]{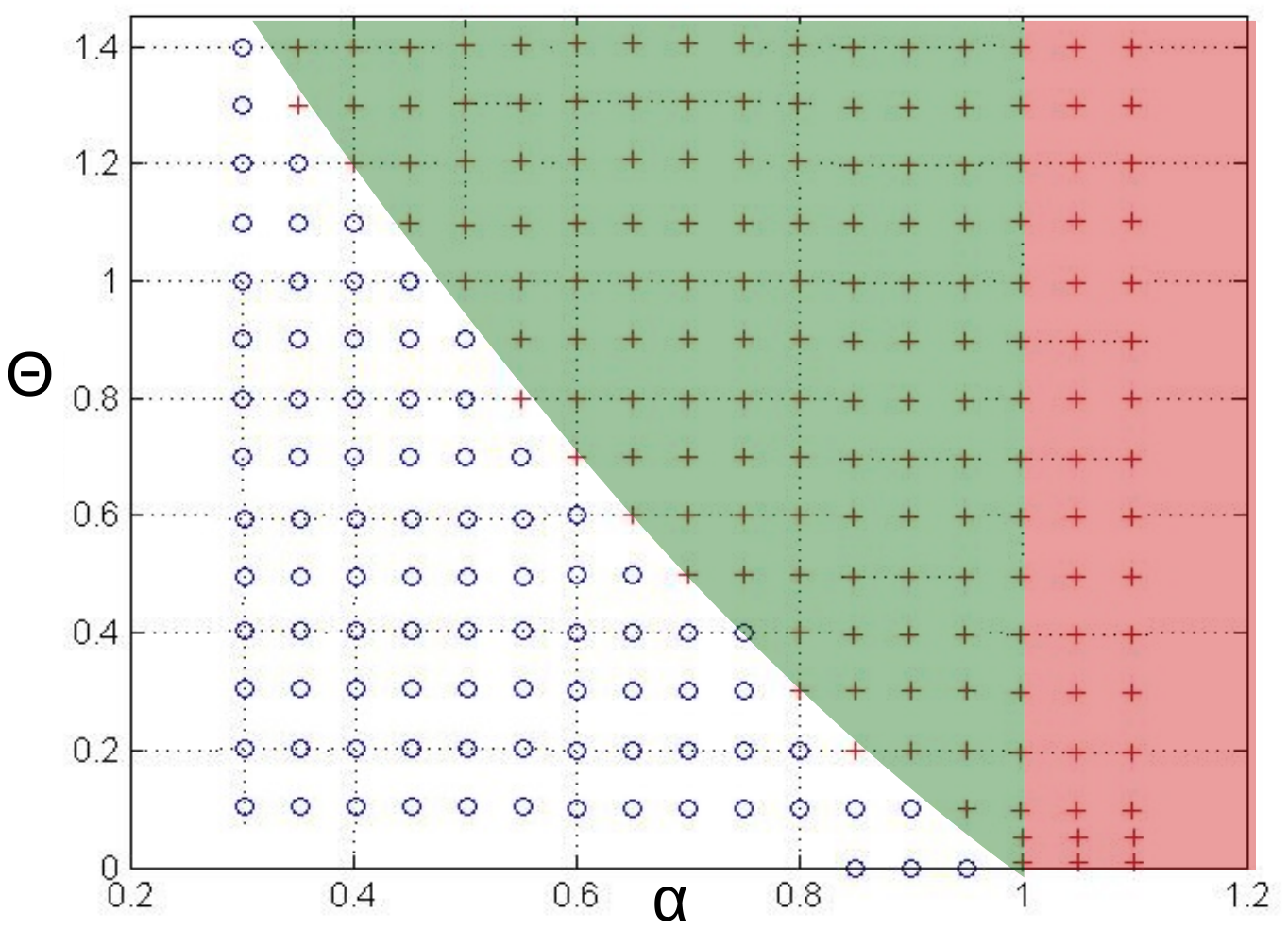}
\caption{{\bf (a)} Maximal stochastic Lyapunov exponent of the 1D chain of neuron-like elements. Crosses and hollow circles denote positive and negative values respectively. The noise level range corresponds to the weak-noise regime in Fig.\ref{Fig___1}. Positive Lyapunov exponents and its vanishing width in the deterministic limit confirm that the N-phase can indeed be identified as noise-induced chaos.}
\label{Fig___3}
\end{figure}

At $\alpha>1$, the vacuum, $x(rt)=0$, looses its stability. This is the onset of the C-phase. We note that in the general case, the loss of stability of a stable vacuum does not necessarily suggest the onset of chaotic behavior. Instability may as well mean a bifurcation. It does, however, indicate the onset of chaotic behavior in our case as can be seen from positive Lyapunov exponents for $\alpha>1$ at zero temperature (Fig.\ref{Fig___3}a). 

At the end of the previous section, the relation between topological entropy and spontaneous TS breaking was discussed. It is also known that topological entropy is related to the Lyapunov exponents via, \emph{e.g.}, the Pesin formula \cite{Pesin} saying that $S$ is a sum of positive Lyapunov exponents. This is actually the very reason why we use the stochastic Lyapunov exponents (see, \emph{e.g.}, Ref.\cite{Arnold1988,Grassberger1988}) as an indication of the spontaneous breakdown of TS. It is likely that more analytical work has to be done before an exact relation between stochastic Lyapunov exponents and spontaneous TS breaking is rigorously established. Nevertheless, as can be seen in Fig.\ref{Fig___3}a, stochastic Lyapunov exponents do reproduce the weak-noise regime of the STS phase diagram in Fig.\ref{Fig___1}. Thus, stochastic Lyapunov exponents in our case support the idea that the N-phase must be identified as noise-induced chaos.

\begin{figure*}[t]
\centering
\includegraphics[width=0.9 \linewidth]{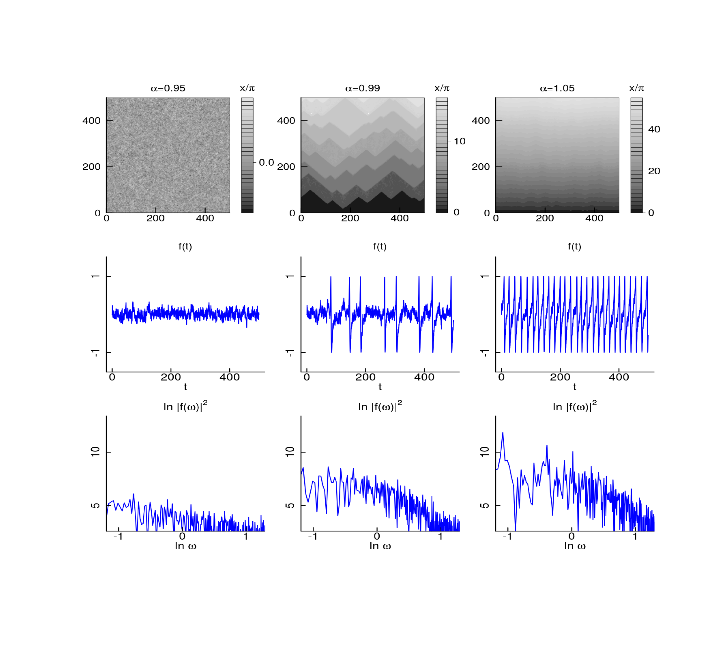}
\caption{ (Top row) Filled contour plots with the horizontal and vertical axes being respectively the time, $t$, and the spatial coordinate, $r$, of simulated $x(rt)$ at a low noise level ($\Theta=0.05$) and three values of $\alpha=(0.95,0.99,1.05)$ from the three dynamical phases as seen from Fig.\ref{Fig___3}. Dynamics in the N-phase (middle) is clearly dominated by anti-instantonic processes (\emph{c.f.}, Fig.\ref{Fig___2}d). The middle and bottom rows are respectively $f(t)=\sin x(r_0t)$ at some position $r_0$ as a function of time and the corresponding power-spectrum. The data reveals the qualitative difference between dynamics in the three major phases and it has the same features as the data from the emulated dynamics in neuromorphic hardware in Fig.\ref{Fig___5}b.}
\label{Fig___4}
\end{figure*}

Yet another important result in this section is the demonstration of the instantoic character of stochastic dynamics in the N-phase. As can be seen in Fig.\ref{Fig___4}, the dynamics in the N-phase is qualitatively different from that in the other two major phases (T- and C-) and just as expected (see Fig.\ref{Fig___2}d) is clearly dominated by (anti-)instantoic processes of creation and annihilation of solitonic configurations, the kink and antikinks.

\section{Emulation of stochastic ND}
\label{Emulation}

In this section, we make one more step toward ND -- one of the most interesting and promising potential applications for STS in the future. More specifically, we construct the STS phase diagram using emulation of stochastic ND on  neuromorphic hardware~\cite{Karlheinz1,Karlheinz2}. In a certain sense, the neuromorphic hardware bridges numerical models and real brains and it offers a few advantages over both. Indeed, unlike experiments in the real brain, neuromorphic hardware provides control over many parameters including the intensity of the noise, which is crucial for our purposes as the noise intensity is one of the key parameters of stochastic dynamics. In addition, as compared to the model in the previous section, the neuromorphic hardware allows for any topology of the network as compared to the oversimplified 1D lattice.

As we already stated before, the reason why the simplest possible network topology was used in the previous section is the spatio-temporal structure of instantonic processes -- the predecessors of neuroavalanches -- that revealed the instantonic character of the dynamics in the N-phase. In real neuronal networks, due to the intricate pattern of interneuronal connections, neuronal avalanches do not have a clear spatial structure of a propagating boundary separating postfired from prefired neurons. \footnote{To be more accurate, this spatial structure of neuronal avalanches can be revealed by embedding the neuronal network into a lattice of sufficiently high dimensionality such that all the directly-connected neurons are neighbors on this lattice. For a square lattice, the number of neighbors is $2D$ with $D$ being the dimensionality of the embedding space. Thus, any given network can be embedded into a square lattice with sufficiently high $D$.} This lack of a clear spatial structure of neuroavalanches makes it difficult to study them experimentally. Therefore, in this section, we will use the 1/f noise characteristic to differentiate between the fundamental dynamical phases. We note that in the literature, avalanche statistics is used more often for this particular purpose (see, \emph{e.g.}, Ref.\cite{Levina1}). There is no conflict here. Both, the $1/f$ noise and the scale-free avalanche statistics can be viewed as signatures of the spontaneous TS breaking. In fact, the very concept of SOC and the associated scale-free statistics was originally introduced as an explanation for $1/f$ noise \cite{Bak1}. 

Before we proceed with our emulation results, a few words are in order about the feasibility of the concept of a ND phase diagram. Anesthesiologists use drugs to render the brain transiently unconscious during surgical procedures. From a physicist's point of view, this means that the blood concentration of the anesthetic is an externally controllable parameter that is being used to draw the brain out of the dynamical phase consistent with consciousness. This physical picture of brain activity  
is supported by experiments under anesthesia showing that the collective neuronal behavior changes suddenly at certain transition points as one changes, via the gradual change of the concentration of a pharmacological agent such as isoflurane, the single-neuron parameters such as the resting potential \cite{Hudson24062014,Ries01041999}. This experimentally demonstrates the existence of qualitatively different dynamical phases in the real brain with sharp transitions separating them, which clearly validates the concept of the ND phase diagram. 



\subsection{Emulation results}


In this section we present our results of emulation of ND using the \emph{Spikey} neuromorphic hardware~\cite{Karlheinz1,Karlheinz2}. The chip is a configurable mixed-signal CMOS implementation of 384 leaky-integrate-and-fire (LIF) neurons with a maximum of 256 synapses each. It features short- and long-term synaptic plasticity and operates in an accelerated mode approximately 10,000 times faster than real-time. Each neuron is configured to have a fixed number of pre-synaptic partners that are randomly selected from the set of available neurons. The intensity of the noise is controlled by a parameter representing the average time interval between two consecutive noise stimuli uniformly distributed within the time interval of the emulation (1000 ms). In other words, this parameter is the reciprocal of the noise intensity $\Theta$ introduced in Eq.(\ref{SDE}).

\begin{figure}[h]
\centering
\includegraphics[width=0.6 \linewidth]{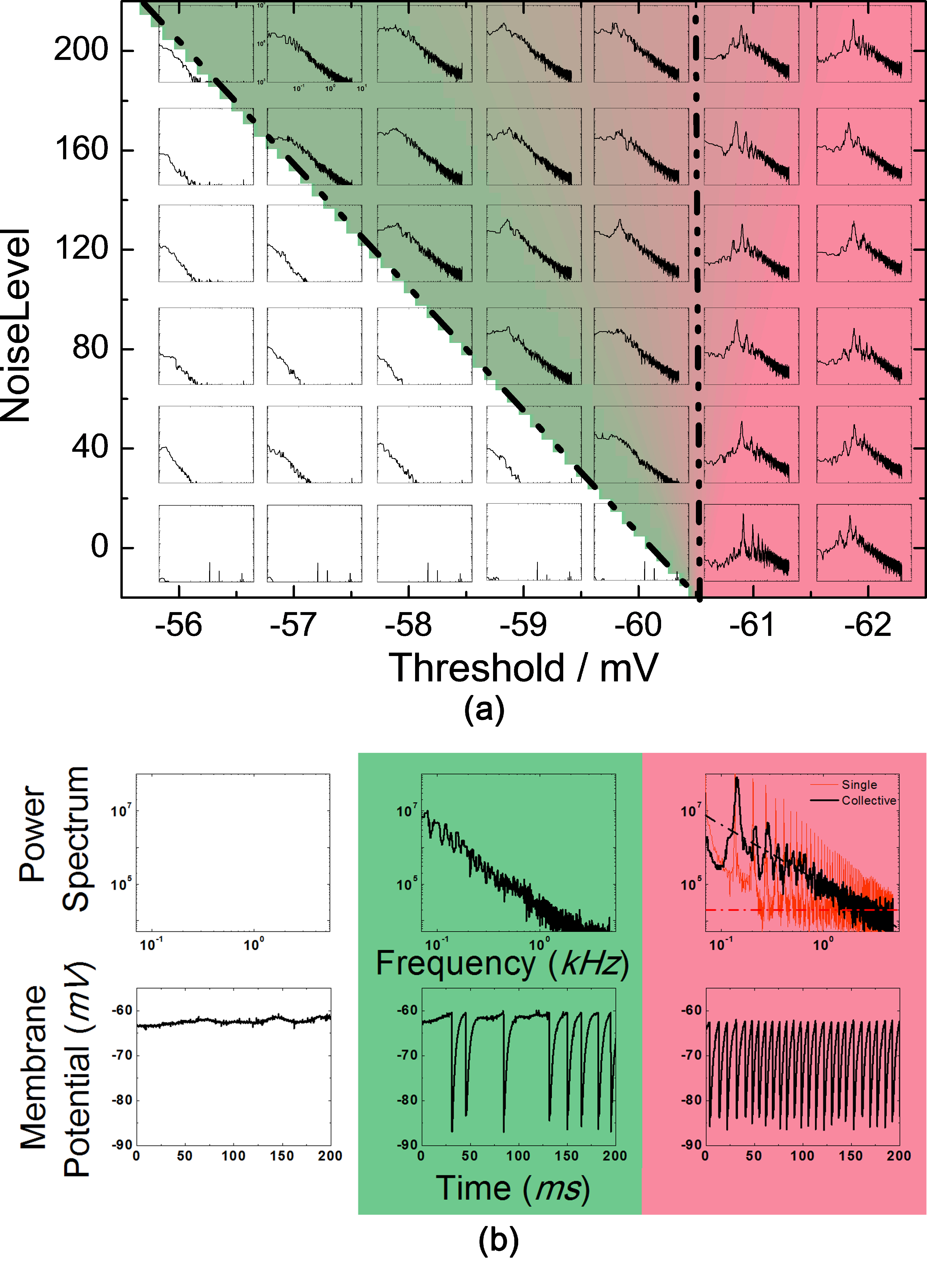}
\caption{{\bf (a)}. The phase diagram of emulated ND using the \emph{Spikey} neuromorphic chip on the plane of the noise intensity and the firing threshold. The insets present the power-spectra of the membrane potential of a neuron and are centered at the points in the phase diagram corresponding to the parameters used for the emulation. The scale of the insets are the same everywhere. The results show that in the deterministic limit, the N-phase collapses onto a sharp transition between the T-phase and C-phase (vertical dashed line at around -60.0 mV), as predicted by the STS picture of the N-phase dynamics; (b). Three typical power-spectra (top) with their corresponding membrane potential recordings (bottom). The thermal equilibrium phase ("coma"-like, the noise-induced chaotic phase ("conscious"-like), and the regular chaotic phase ("seizure"-like) are featured, respectively, by no membrane potential dynamics with a sharp decrease at low frequency on power spectra (left column), avalanche-like membrane potential with 1/f noise–like spectra (middle column), and a non-stop firing pertinent to the seizure-like collective neuronal behavior with 1/f noise-like spectra superimposed by equidistant peaks (right column) representing periodic dynamics. This power-spectrum is also compared to the one generated by an isolated neuron (lower red curve). As can be seen, the spectrum of the isolated neuron does have a 1/f noise substrate, which is thus a signature of the collective neuronal dynamical behavior in the network.}
\label{Fig___5}
\end{figure}

\begin{figure}[h]
\centering
\includegraphics[width=0.6 \linewidth]{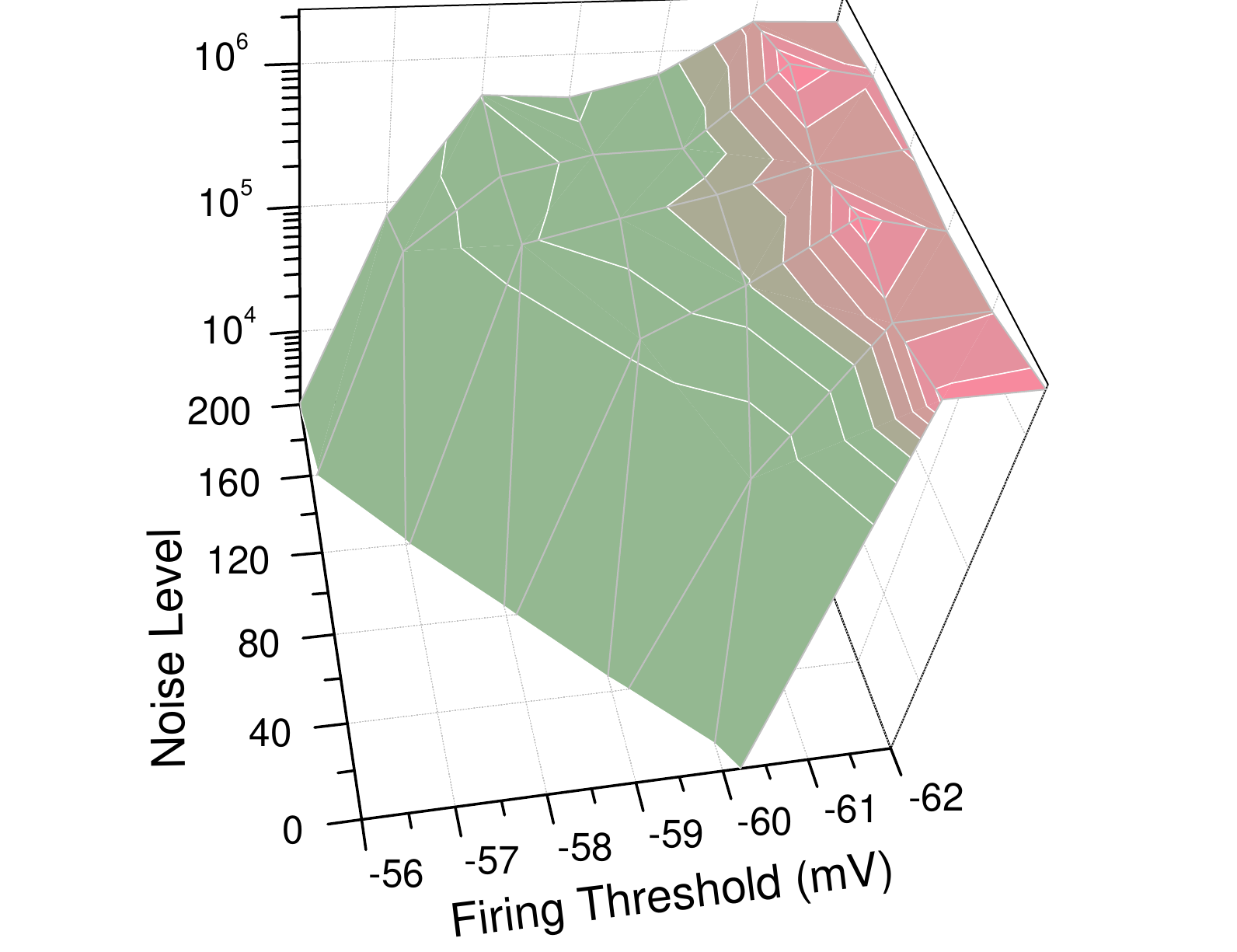}
\caption{{\bf (a)}. The integral intensity of dynamics defined as $\int |f(\omega)|^2 d\omega$, where $f(\omega)$ is the Fourier component of the membrane potential of a neuron. Even though this characteristic cannot be viewed as the "chaotic" order parameter, it does experience a relatively sudden jump and a smeared plateau at, respectively, the T-N and the N-C transition lines of Fig.\ref{Fig___5}.}
\label{Fig___6}
\end{figure}

In Fig.\ref{Fig___5}, the power-spectra of a membrane potential of one neuron are given for different firing threshold potentials and noise intensities and in Fig.\ref{Fig___6}, we plot the characteristic representing the overall "intensity" of dynamics, $\int |f(\omega)|^2 d\omega$; here $f(\omega)$ is the Fourier component of the membrane potential. As can be seen, the results clearly reproduce the three-part phase diagram: the subthreshold T-phase with no conspicuous dynamics, the N-phase, featured by a $1/f$ power spectrum, and the C-phase with the $1/f$ power spectrum superimposed with equidistant peaks reflecting the approximate time-periodicity of permanent firing above threshold.

It must be noted that, just like in the real brain, the topology of neuronal network is random and neuroavalanches do not have a well-defined spatial structure similar to the spatial propagation of solitons discussed in the previous section. Therefore, revealing this structure is not among our goals in this section. Our main objective here is to reveal that the N-phase collapses onto the border of the C-phase in the deterministic limit. This objective is the reason why the ND emulation is conducted over a broad range of noise intensity. 

We would also like to point out that it was known previously that the presence of noise often leads to the emergence of power-laws in dynamical systems near the border of chaotic activity and this fact was proposed to be viewed as the reason behind power-laws in ND \cite{ThresholdProcesses}. The STS picture of the N-phase dynamics provides a solid theoretical explanation why this happens. 

In the real brain, however, the intensity of the noise is not an externally controllable parameter unless, of course, one views the sensory input also as a part of the noise. We find it more physically appealing, however, to view stimulii as a perturbation of the ground state of the brain, and the response to this perturbation is the essence of information processing within ND. Thus, the noise in our approach represents the (intractable) bio-chemo-electric influence from the host body only. Therefore, the intensity of the noise is not an externally controllable parameter of the ND in the real brain. 

\begin{figure}[h]
\centering
\includegraphics[width=0.6 \linewidth]{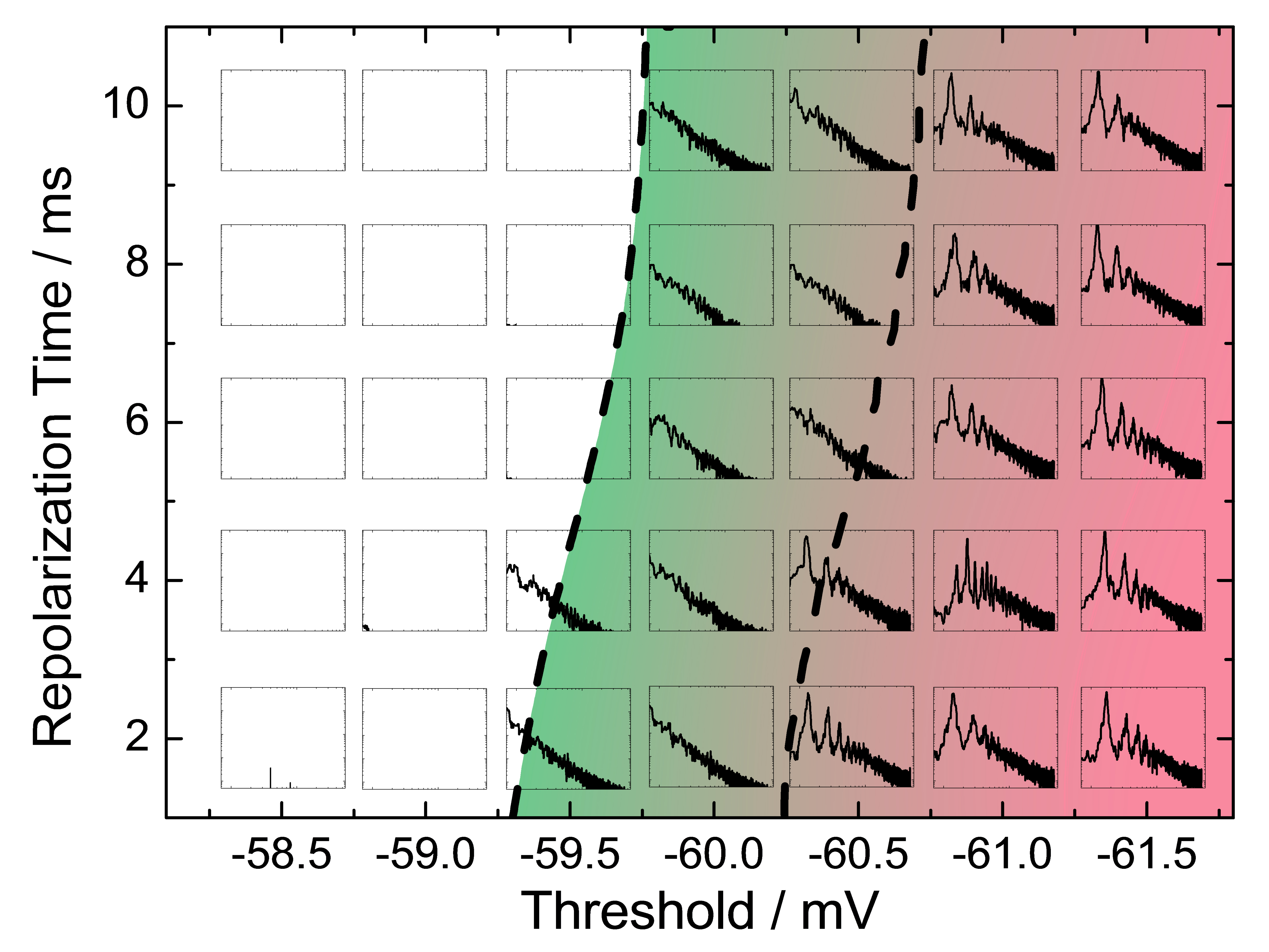}
\caption{The phase diagram of emulated ND using the \emph{Spikey} chip at a fixed noise intensity of 10 Arb.Units in Fig.\ref{Fig___5} and for various values of the firing threshold and repolarization time. As in Fig.\ref{Fig___5}, the insets show the power spectra at the corresponding values of the parameters. As expected, the N-phase, featured by the $1/f$-type spectra, is sandwiched between the T- and the C-phases. The results show that the position of the N-phase shifts to the right with the increase of the repolarization time, thus pointing out that the effect of the increase of the repolarization time is similar to the increase of the firing threshold.}
\label{Fig___7}
\end{figure}

At the same time, there are many other externally controllable single-neuron parameters relevant to the ND in the brain. One of the parameters is the neuron repolarization time. In fact, we believe that, to a good approximation, the firing threshold and the repolarization time are the two major externally controllable parameters of the collective ND in the real brain. For this reason, we also constructed the phase diagram of the emulated ND on the plane of these two parameters. As is expected and seen in Fig.\ref{Fig___7}, at fixed noise intensity, the N-phase is sandwiched between the T- and the C-phases.

\subsection{Neurodynamic meaning of the three phases}


The TS breaking picture of chaotic dynamics puts the phenomenon of chaotic dynamics in the brain into a new and promising perspective. The point is that the spontaneous breakdown of various symmetries is known in physics as ''ordering'' (crystalline order, ferromagnetic order, \emph{etc.}). Therefore, the true essence of chaotic dynamics is quite opposite to semantics of the word "chaos" and the common perception of this phenomenon. Unlike "chaos", this new understanding, which can be dubbed the dynamical long-range order (DLRO), has a positive rather than negative connotation in the context of information processing in the brain and it allows one to make a first step in what seems to be a new view on the functionality of brain activity.   

Namely, in behavioral sciences and psychology, there is a concept of a short-term memory operating on the order of seconds. We find it natural to believe that the short term memory must be directly connected or rather based on the spontaneous ND memory associated with the spontaneous TS breaking. After all, why should nature not take advantage of the fact that DLRO provides effortless long-range dynamical information storage within the brain for short-term memory purposes ? This argument is especially convincing in light of the fact that this advantage comes at no extra cost from the bio-chemical point of view, \emph{i.e.}, there is no need to invoke some specialized bio-chemical process(es) that would be responsible for it.

The reasoning in the previous paragraph suggests that a conscious brain can reside only in either the N- or C- phases, where the TS is spontaneously broken and/or the DLRO is present. Given that, the non-stop firing of neurons in the C-phase is very reminiscent of the ND phenomenon of epileptic seizure \cite{seizure}, therefore, the possibility that a conscious brain is in the C-phase can be ruled out. As a result, one is left with the conclusion that a conscious brain can reside only in the N-phase, which thus can be identified as a ''conscious-like'' phase. Accordingly, the T- and the C- phases can be dubbed as the ''coma-like''\footnote{Note that coma is a behavioral state of unresponsiveness, not a brain state, like seizure. That is, one can have localized seizures in the brain that result in a coma. Therefore, “quiescent”-like may be a better identifier for the T-phase of ND. Nevertheless, We will use “coma-like” phase as in the everyday language; "coma" reflects better the characteristics of the T-phase.} and the ''seizure-like'' phases, respectively.

We find it likely that somewhere on the right hand side of the ND phase diagram, \emph{i.e.}, on the side of the high resting potentials well above the firing threshold, there may exist a phase of synchronized persistent neuronal oscillations. The transition from this phase to the C-phase would correspond to what is known in the literature as the period doubling route to chaos. This transition may be an interesting object of future studies.

We would like to point out that the importance of noise for healthy brain operation has been discussed previously \cite{Beck201230,Deco20091,Rolls2012212}. In our case, this importance is taken to the extreme -- the "conscious-like" phase does'nt even exist without the noise.

\begin{figure}[t]
\centering
\includegraphics[width=0.6 \linewidth]{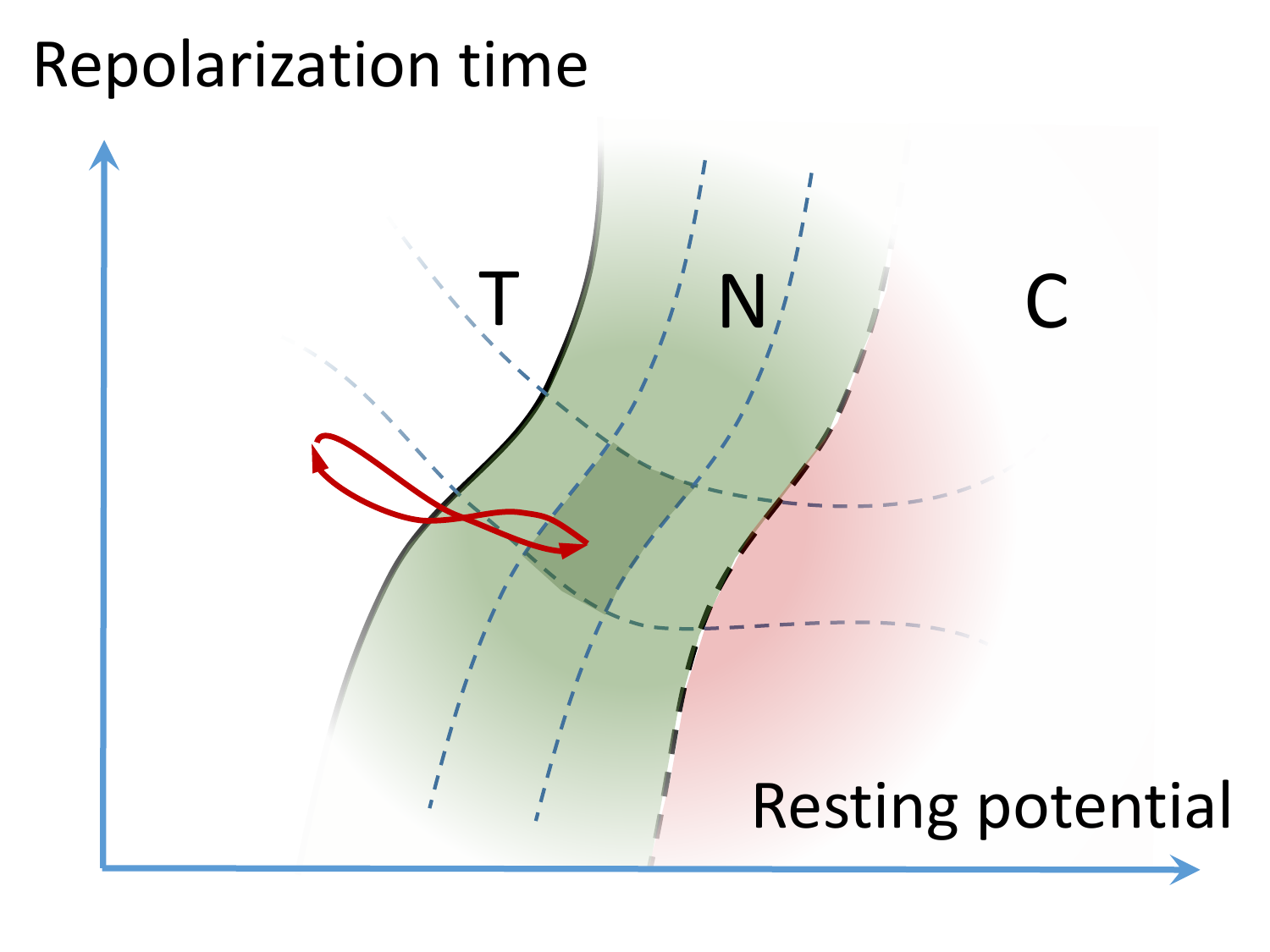}
\caption{A hypothetical phase diagram of the ND with the rudimentary structure as displayed in Fig. \ref{Fig___7} and with an additional fine-structure that resolves the larger scale collective dynamical phenomena (see discussion in text). The boundaries between the qualitatively different subphases are presented as dashed curves. The "awakeness" phase (the thicker green filled rectangle-like area) must be a subphase of the N-phase. The two closed red arrows qualitatively represent the phase space trajectory during an anestheological cycle.}
\label{Fig___8}
\end{figure}

\subsection{Fine structure}

The phase diagram in this paper is very rudimentary, which is a fair price for its generality. It is understood that the ND phase diagram of a real brain must have a fine-structure on top of the three-phase picture discussed here (see Fig.\ref{Fig___8}). This fine structure must resolve various types of interactions between different parts of a real inhomogeneous brain \footnote{For example, the ''real conscious'' phase and the ''sleeping phase'' may as well be two different neighboring subphases of the N-phase; the subphases that are separated by a phase-transition-like boundary.}. Even though our understanding of the fine-structure of the ND phase diagram is qualitative, it may be already possible to use this concept for visualization of the effect of a pharmacological agent on the collective ND during a typical anesthesiological cycle. As the blood concentration of the agent gradually changes, the ND moves on the phase diagram and changes qualitatively as it occasionally crosses the critical boundaries between subphases.\cite{Hudson24062014}

\section{Conclusion}
\label{Conclud}

The recently proposed approximation-free supersymmetric theory of stochastics, spontaneous breakdown of topological supersymmetry that all stochastic differential equations possess is the stochastic generalization of the concept of dynamical chaos. 
The theory also offered an explanation for the pre-chaotic phase known previously as self-organized criticality. As it turns out, at non-zero noises, ordinary chaos is preceded by a noise-induced chaotic phase where the topological supersymmetry is broken by the condensation of noise-induced (anti)instantonic configurations. In this paper, we supported this picture by numerical studies of a 1D chain of neuron-like elements and experimental emulation of stochastic neurodynamics using neuromorphic hardware. We demonstrated that the stochastic Lyapunov exponents are positive in the N-phase, the dynamics is indeed dominated by the (anti)instantnoic processes, and the ''width'' of the N-phase vanishes in the deterministic limit.

The novel supersymmetry breaking understanding of chaotic dynamics suggests that studies of the corresponding spontaneous dynamical order in neurodynamics can be the right venue toward understanding the high level functionalities and the fundamental principles of information processing in the brain. More specifically, one of the most fruitful directions for further research could be the development of low-energy effective theories (LEET) for N-phases. Unlike in the Wilson-Cowan \cite{WILSON19721} and others approaches to neurodynamics (for a review, see, \emph{e.g.}, Ref.\cite{WSBM:WSBM1348}), the LEET for the spontaneous dynamical order must be written for a fermionic field, which must be the order parameter for the spontaneously broken fermionic symmetry. Although at this moment we do not know much about the details of such an LEET, we find it reasonable to believe that if a computational paradigm at least remotely related to the principles of the information processing/encoding by neurodynamics already exists, it is most likely the topological or fault-tolerant quantum computing \cite{TopQuant}. After all, the supersymemtric theory of stochastics is a rightful member of the family of the Witten-type topological field theories.

We would like to conclude our discussion by pointing out that the application of supersymemtric theory of stochastics to neurodynamics bridges mathematical physics and anesthesiology and further work in this direction may result in fruitful cross-fertilization between science and medicine.





\section*{Acknowledgments}
K.L.W. would like to acknowledge the support of the endowed Raytheon professorship. A.E.H. would like to acknowledge the support of the Foundation for Anesthesia Education and Research Mentored Research Training Grant. The neuromorphic hardware and software is partially supported by EU Grant 269921 (BrainScaleS), and EU Grant 604102 (Human Brain Project, HBP). We would like to thank Thomas Pfeil for his support of the neuromorphic hardware system.


\bibliographystyle{elsarticle-num-names}
\bibliography{Crit_vs_SUSY}

\end{document}